\documentclass[11pt]{article}
\usepackage{amssymb,amsmath,latexsym,slashed}
\newcommand{\ft}[2]{{\textstyle\frac{#1}{#2}}}

\def\rmi{{\rm i}}
\def\rme{{\rm e}}
\newcommand{\SU}{\mathop{\rm SU}}
\newcommand{\SO}{\mathop{\rm SO}}
\newcommand{\U}{\mathop{\rm {}U}}

\newcommand{\OSp}{\mathop{\rm {}OSp}}

\newcommand{\Sl}{\mathop{\rm {}S}\ell }

\usepackage{ifpdf}
\ifpdf
   \pdfoutput=1
   \pdftrue
  \usepackage[pdftex]{hyperref}
 \else
   \usepackage{cite}
\fi

\makeatletter
\@addtoreset{equation}{section}
\makeatother

\def\bftau{\mbox{\boldmath $\tau$}}

  \let\g=\gamma  \let\e=\epsilon

\let\m=\mu \let\n=\nu

\newcommand{\be}{\begin{equation}}
\newcommand{\ee}{\end{equation}}

\begin{document}

\begin{titlepage}
\rightline{arXiv:0704.3918 [hep-th]}
\rightline{DAMTP-2007-40, ITFA-2007-16, KUL-TF-07/09}

\vfill

\begin{center}
\baselineskip=16pt
{\Large\bf Domain-wall/Cosmology correspondence \\
\vskip 0.3cm
in adS/dS supergravity}
{\large {\sl }}
\vskip 10.mm
{\bf ~Kostas Skenderis$^{*,1}$, Paul K. Townsend$^{\dagger,2}$
and  Antoine Van Proeyen$^{\star,3}$}
\vskip 1cm
{\small
$^*$
Institute for Theoretical Physics,
University of Amsterdam,\\
Valckenierstraat 65, 1018 XE Amsterdam,
The Netherlands\\
}
\vspace{6pt}
{\small
$^\dagger$
Department of Applied Mathematics and Theoretical Physics,\\
Centre for Mathematical Sciences, University of Cambridge,\\
Wilberforce Road, Cambridge, CB3 0WA, U.K. \\
}
\vspace{6pt}
{\small
$^\star$
Instituut voor Theoretische Fysica, Katholieke Universiteit Leuven,\\
       Celestijnenlaan 200D, B-3001 Leuven, Belgium.}
\end{center}
\vfill

\par
\begin{center}
{\bf
ABSTRACT}
\end{center}
\begin{quote}

We realize the domain-wall/cosmology correspondence for
(pseudo)supersymmetric domain walls (cosmologies) in the context of
four-dimensional supergravity.  The $\OSp(2|4)$-invariant anti-de Sitter
(adS) vacuum of a particular $N=2$ Maxwell-Einstein supergravity theory
is shown to correspond to the $\OSp(2^*|2,2)$-invariant  de Sitter (dS)
vacuum of a particular pseudo-supergravity model, with `twisted' reality
conditions on spinors. More generally, supersymmetric domain walls of the
former model correspond to pseudo-supersymmetric cosmologies of the
latter model, with time-dependent pseudo-Killing spinors that we give
explicitly.

\vfill

\vfill

 \hrule width 5.cm
\vskip
2.mm
{\small
\noindent $^1$ skenderi@science.uva.nl \\
\noindent $^2$ p.k.townsend@damtp.cam.ac.uk \\
\noindent $^3$ antoine.vanproeyen@fys.kuleuven.be }
\end{quote}
\end{titlepage}

\newpage

\section{Introduction}

A scalar field minimally coupled to gravity is said to define a `fake supergravity'  theory \cite{Freedman:2003ax}  if the scalar potential $V$ is given in terms of a triplet superpotential
${\bf W}$ by a certain `supergravity-inspired'  formula (see  \cite{Skenderis:1999mm,DeWolfe:1999cp} for related earlier work, and \cite{Celi:2004st,Zagermann:2004ac,Sonner:2007cp,Elvang:2007ba} for  recent discussion of the multi-scalar, and other, generalizations).  A domain-wall solution supported by the scalar field is then said to be (fake) supersymmetric if it admits a `Killing spinor', defined as a non-zero solution for the complex doublet spinor field $\chi$ of the `Killing spinor' equation
\be\label{KillingSpinor}
\left(D_\mu + {\bf W} \cdot \bftau \, \Gamma_\mu\right)\chi=0\, , \qquad (\mu=0,1,\dots,D-1)\,,
\ee
where $D_\mu$ is the standard covariant derivative acting on Lorentz spinors, $\Gamma_\mu$ are Dirac matrices  and $\bftau$ is a triplet of Pauli matrices. Remarkably, all  domain walls that are
either flat or  `adS-sliced' (foliated by anti-de Sitter spacetimes) are (fake) supersymmetric if the
scalar field  is strictly monotonic,  because under these circumstances the required triplet superpotential can be constructed from the solution itself \cite{Freedman:2003ax,Sonner:2005sj, Skenderis:2006jq,Skenderis:2006rr}.  The superpotential so constructed turns out to be real, and it takes the form ${\bf g}W$ for a flat wall, where  ${\bf g}$ is a {\it fixed} 3-vector and $W$ a real scalar superpotential.

It was shown in \cite{Skenderis:2006jq,Skenderis:2006fb} that similar results apply to
flat and closed homogeneous and isotropic (Friedmann-Lemaitre-Robinson-Walker (FLRW)) cosmologies, despite their time-dependence.
This result was found by an application of a `domain-wall/cosmology (DW/C)
correspondence', which states that for every maximally symmetric domain-wall solution of
gravity coupled to scalar fields of a model with potential $V$ there is a corresponding homogeneous
and isotropic cosmology of the same model but with potential $-V$ \cite{Cvetic:1994ya,Skenderis:2006jq}. If  the domain wall is adS-sliced then the corresponding cosmology is closed, and if the domain wall is flat then so is the corresponding cosmology. In either case, a domain-wall solution that is fake supersymmetric with respect to a (real) superpotential ${\bf W}$ corresponds to a cosmology that is fake supersymmetric with respect to the imaginary superpotential $\rmi{\bf W}$. The
corresponding solution of (\ref{KillingSpinor}) was called a ``pseudo-Killing'' spinor in \cite{Skenderis:2006jq,Skenderis:2006fb} because the `gamma-trace' of this equation is a Dirac-type equation but with an {\it anti-hermitian}  `mass' matrix.  In this paper we will need only the special case of this result for flat domain walls and the corresponding flat cosmologies. Each fake supersymmetric flat domain wall is paired with a fake supersymmetric flat cosmology, and this pair is associated with some real scalar superpotential $W$ such that there exist non-zero solutions $\chi$ of the equation
\be\label{KSflat}
\left(D_\mu + W\,  {\bf g} \cdot \bftau \, \Gamma_\mu\right)\chi=0\, ,
\ee
for some fixed 3-vector ${\bf g}$ that is real for the domain wall and imaginary for the cosmology.

Just as some fake supersymmetric domain-wall solutions of a fake
supergravity theory may also be `genuinely' supersymmetric solutions of
`genuine' supergravity theory (for a restricted set
of possible spacetime dimensions $D$), one might expect some `fake'
pseudo-supersymmetric cosmological solutions to be `genuinely'
pseudo-supersymmetric solutions of  some `genuine' supergravity theory.
However, Killing spinors arising in supergravity theories are generally subject to  some reality (and/or chirality) condition. For example, for $D=5$ the Killing spinor equation (\ref{KillingSpinor}) can be deduced from the condition of  vanishing supersymmetry variation of the gravitino field of $D=5$ supergravity coupled to matter  \cite{Celi:2004st}, and in this context the spinor $\chi$ is subject to a symplectic-Majorana condition that requires ${\bf W}$ to be real. Similar considerations apply in other dimensions and are expected to lead to the same conclusion (although the precise relation of the `fake' Killing spinor equation (\ref{KillingSpinor}) to the supergravity  supersymmetry preservation conditions is known in only a few cases).

However, there exist  `non-standard' supergravity theories that are found by imposing `twisted reality' conditions on spinors; we shall call them pseudo-supergravity theories. They first arose from an investigation of whether there could be supergravity theories with supersymmetric de Sitter (dS) vacua. The dS supergroups available as isometry supergroups were classified by Nahm \cite{Nahm:1977tg}
and are listed in table~\ref{tbl:superdS}, along with the R-symmetry
group.
 \begin{table}[ht]
  \caption{\it de Sitter supergroups in $D\geq 4$ whose bosonic subgroup is $\SO(D,1)\times R$.}
  \label{tbl:superdS}
\begin{center}
  \begin{tabular}{|c|l|rl|}
\hline
  & supergroup & \multicolumn{2}{c|}{R-symmetry} \\
\hline
$D=6$  & F${}^1(4)$ &  & $\SU(2)$ \\
\hline
$D=5$  & $\SU^*(4|2n)$ & $n=1$ & $\SO(1,1)\times \SU(2)$ \\
  &  & $n=2$ & $\SO(5,1)$ \\
\hline
$D=4$  & $\OSp(N^*|2,2)$ & $N=2$ & $\SO(2)$ \\
  &  & $N=4$ & $\SU(1,1)\times  \SU(2)$ \\
  &  & $N=6$ & $\SU(3,1)$ \\
  &  & $N=8$ & $\SO(6,2)$ \\
\hline
\end{tabular}
\end{center}
\end{table}
Applications for $D=4,5,6$ have been discussed in
\cite{Lukierski:1984,Pilch:1984aw,Lukierski:1984it,deWit:1987sn,D'Auria:2002fh}; in
particular, an explicit $N=2$, $D=4$ `de Sitter' pseudo-supergravity was
constructed in \cite{Pilch:1984aw,deWit:1987sn}. Just as adS space can be
viewed as a special case of a domain wall, so dS space can be viewed as a
special case of an FLRW cosmology. This suggests that it should be
possible to view pseudo-supergravity theories with supersymmetric dS
vacua as `duals' of `standard' supergravity theories with supersymmetric
adS vacua, such that these two vacua are `dual' in the sense of the DW/C
correspondence.  One purpose of this paper is to confirm this logic for a
particular $N=2$, $D=4$, $\U(1)$-gauged Maxwell-Einstein
pseudo-supergravity theory that we show to be `dual' in the sense just
described to a `standard'  $N=2$, $D=4$ Maxwell-Einstein supergravity.

A further purpose of this paper is to extend this result to generic
(pseudo)supersymmetric domain walls (cosmologies) of these `dual'
theories, although we shall limit ourselves here to solutions that are
asymptotic to the adS (dS) vacuum.  In either case the
(pseudo)supersymmetry is shown to be a consequence of the existence of a
non-zero solution for $\chi$ of  (\ref{KSflat})  with ${\bf g}=(g,0,0)$,
where $g$ is the gauge coupling constant. While the standard reality
conditions on $\chi$ imply that $g$ is real, the twisted reality
conditions imply that it is imaginary, exactly as required  for
pseudo-supersymmetry. When $g$ is imaginary, reality of the action
requires the $\U(1)$ gauge field to  be imaginary too. This means that
the kinetic term for a redefined, real, gauge field is negative, so the
pseudo-supergravity has vector ghosts \cite{Pilch:1984aw,deWit:1987sn};
this is a manifestation of the fact that  there is no non-trivial
representation of a  dS superalgebra in a positive definite Hilbert
space.

This paper is organized as follows. In the next section we recall  the
essentials of $N=2$, $D=4$  supergravity and the different reality
conditions that one may impose on the spinors. In particular, for a
choice of gauging we show that standard reality conditions lead to a
supersymmetric adS critical point while twisted reality conditions to a
supersymmetric dS critical point. We further relate this to fake
supergravity. In section \ref{ss:DWCcorr} we realize the
domain-wall/cosmology correspondence in supergravity by finding the
corresponding supersymmetric domain-wall/cosmology solutions. Section
\ref{ss:disc} contains our conclusions and further discusses some
implications.

We became aware of related work on a realization of the DW/C
correspondence in $D=10$ and $D=11$ supergravity \cite{Bergshoeff:2007cg}
during the completion of an earlier version of this paper. This revision
presents the precise relation of our supergravity results to the fake
supergravity formalism in which context the correspondence was originally
proposed.

\section{$N=2$ gauged (pseudo)supergravity}
 \label{ss:N2gauged}

In this section we review the features of $D=4$, $N=2$ Einstein-Maxwell
supergravity theory \cite{deWit:1984pk} relevant for our application, in
particular the choices of reality conditions on spinors, and we also show
how the (pseudo)Killing spinor  equation (\ref{KSflat}) arises in this
context. For one vector multiplet, the bosonic fields are the metric
$G_{\m \n}$, two vectors $A_\m^I\ (I{=}0,1),$ (with $A^0_\m$ being the
graviphoton) and a complex scalar $z$; the fermionic fields are the two
gravitini $\psi^i_\mu$ and the two photini $\lambda^i$, ($i=1,2$). A
linear combination $g_IA^I_\m$ of the two vector fields may be used to
gauge  a $\U(1)$ group. We will consider a model with an
$\SO(1,1)$-invariant metric on the  space of coupling vectors $g_I$,
leading to three types of gauging according to whether this vector is
`timelike', `spacelike' or `null'. For the most part we follow the
notations and conventions in \cite{VanProeyen:1999ni}.

\subsection{$N=2$ supergravity with one vector multiplet}

As we are interested in domain-wall/cosmology solutions that involve only
the metric and the scalar fields, we truncate the supergravity theory to
this sector. The truncated Lagrangian density ${\cal L}$  is given by
\begin{equation}
  e^{-1}{\cal L}=\ft12 R
-g_{z \bar z}D_\mu z D^\mu \bar z \ -\ V(z,\bar z)\,,
 \label{Lbos}
\end{equation}
where $e=\sqrt{-\det G}$, and $g_{z\bar z}$ is the K{\"a}hler target space metric, given in terms of a K{\"a}hler potential ${\cal K}$ by
\be\label{targetmetric}
g_{z\bar z} = \partial_z\partial_{\bar z}{\cal K}\, .
\ee
The relations of special geometry imply that
\begin{equation}
  \rme^{-{\cal K}} = -\rmi \left(Z^I \frac{\partial \bar F}{\partial
  \bar Z^I} - \bar Z^I \frac{\partial F}{\partial
  Z^I} \right)\,,
 \label{generalK}
\end{equation}
where $Z^I(z)$ ($I=0,1$) are two holomorphic functions of $z$, and $F(Z)$ is a holomorphic function of these two variables, homogeneous of second
degree. Due to the homogeneity, one of the variables $Z^I$ is irrelevant, so one can take an arbitrary parametrization
(up to some requirements of non-degeneracy) of the $Z^I$ in terms of $z$.

The form of the potential $V$ is determined, via a Ward identity, from
the supersymmetry transformations rules of the gravitini  and  photini
\cite{Cecotti:1985mx,Ferrara:1985gj,Cecotti:1984wn}, so we consider these
first. We take the spinor parameters to be  $\epsilon^i$ ($i=1,2$) and we
work with chiral spinors,  i.e. eigenspinors of $\gamma_5 \equiv
\rmi\gamma_0\gamma_1\gamma_2\gamma_3$,  with the position of the index
indicating chirality:
\begin{eqnarray}
&&\epsilon ^i=+\gamma _5\epsilon ^i\,,\qquad \psi _\mu{}^i=+\gamma _5\psi
_\mu {}^i\,,\qquad \lambda ^i=-\gamma _5\lambda ^i\,,\nonumber\\
&&\epsilon _i=-\gamma _5\epsilon _i\,,\qquad \psi _{\mu
i}=-\gamma_5\psi_{\mu i} \,,\qquad \lambda _i=+\gamma _5\lambda _i\,.
 \label{chirality}
\end{eqnarray}
After truncation to the metric-scalar sector, the fermion field supersymmetry transformation laws are
\begin{eqnarray}
 \delta \psi _{\mu i} &=& \left(D_\mu -\ft12\rmi A_\mu \right) \epsilon_i
 -\gamma_\mu S_{ij}\epsilon^j\,, \qquad
  \delta \lambda ^ i = \slashed{\partial }\bar z \epsilon^i +N^{\bar z\, ij}\epsilon_j\, ,\nonumber \\
 \delta\psi_\mu^i &=& \left(D_\mu + \ft12\rmi  A_\mu \right) \epsilon^i
  -\gamma_\mu S^{ij}\epsilon_j\,, \qquad
\delta \lambda _ i = \slashed{\partial }z \epsilon_i +N^z_{ij}\epsilon^j\,,
 \label{relevsusytransf}
\end{eqnarray}
where $D_\mu$ is the usual Lorentz-covariant derivative on spinors, and
\begin{equation}
  A_\mu =-\ft12\rmi \left( \partial_\mu  z\, \partial_z {\cal K}
  -\partial _\mu  \bar z\, \partial _{\bar z }{\cal K} \right)
 \label{valueAmu}
\end{equation}
is the K{\"a}hler connection. The auxiliary fields are given by
\begin{eqnarray}
  S_{ij} &=& -P_{Iij}\, \rme^{{\cal K}/2} \bar Z^I\,,\qquad
  N^{\bar z\, ij}=  -2\rme^{{\cal K}/2} P_I^{ij}\, g^{z\bar z}{\cal D}_z Z^I\, , \nonumber\\
  S^{ij} &=& - P_I^{ij} \rme^{{\cal K}/2} Z^I\, , \qquad
  N^z_{ij}=  -2\rme^{{\cal K}/2} P_{I ij}
  \, g^{z\bar z}\,\overline{{\cal D}_zZ^I}\, ,
 \label{defSN}
\end{eqnarray}
where $g^{z\bar z}= (g_{z\bar z})^{-1}$, and
\begin{equation}
{\cal D}_z Z^I =  \partial_z Z^I + Z^I \partial_z{\cal K}
\end{equation}
is the K{\"a}hler-covariant derivative of $Z^I$. One also has
\begin{equation}
  P_I^{ij}=\varepsilon ^{ik}\varepsilon ^{j\ell}P_{Ik\ell }\,,
 \label{relationsepsP}
\end{equation}
where $P_{Iij}$ (the moment map) will be specified below. We use
conventions for which $ \varepsilon ^{ij}\varepsilon _{kj}=\delta ^i_k$
and $\varepsilon _{12}=\varepsilon^{12}=1$.
As mentioned, the scalar potential follows directly from the transformation laws, and is given by
\be
V= -6S^{ij}S_{ij}+ \ft12 g_{z\bar z}
N^z_{ij} N^{\bar z ij}\,.
\ee

In the absence of physical hypermultiplets, $P_{Iij}$ are the entries of
two constant symmetric matrices $P_I$, and an `equivariance condition'
requires these matrices to be proportional, so
\begin{equation}
  P_{Iij}= g_I e_{ij}\,,
 \label{gPI}
\end{equation}
for constants $g_I$ (which are the components of the coupling  constant
vector mentioned earlier)  and constant symmetric matrix $e_{ij}$. Introducing
a triplet of Pauli matrices ${\bf \tau}$, we may write
\be \label{triplet}
e_{ij} = \left[ \tau_2\left({\bf n}\cdot {\bf \tau}\right)\right]_{ij}\,,
\ee
for 3-vector ${\bf n}=(n^1,n^2,n^3)$, which is complex, a priori, but will be restricted by  reality conditions to be discussed below.  The  supersymmetry transformations (\ref{relevsusytransf})
may now be written as
\begin{eqnarray}\label{susytrans1}
\delta\psi_{\m i} &=& \left(D_\mu -\ft12\rmi  A_\mu \right) \epsilon_i +
 \rme^{{\cal K}/2} \bar {\cal Z}\gamma_\mu
\left[\tau_2\left({\bf n}\cdot\bftau\right)\right]_{ij}\epsilon^j \, , \nonumber\\
\delta\psi_\m^i &=& \left(D_\mu +\ft12\rmi  A_\mu \right) \epsilon^i +
\rme^{{\cal K}/2} {\cal Z}\gamma_\mu
\left[\left({\bf n}\cdot\bftau\right)\tau_2\right]^{ij}\epsilon_j \, , \nonumber\\
\delta \lambda^ i &=& \slashed{\partial }\bar z \, \epsilon^i -2\rme^{{\cal K}/2}g^{z\bar z}\,  {\cal D}_z {\cal Z}
\left[\left({\bf n}\cdot\bftau\right)\tau_2\right]^{ij}\epsilon_j \, , \nonumber\\
\delta\lambda_i &=& \slashed{\partial }z\,  \epsilon_i -2\rme^{{\cal K}/2}g^{z\bar z}\, \overline{{\cal D}_z{\cal Z}}
\left[\tau_2\left({\bf n}\cdot\bftau\right)\right]_{ij}\epsilon^j \, ,
\end{eqnarray}
where
\begin{eqnarray}
 &&{\cal Z} = g_IZ^I \, , \qquad
{\cal D}_z {\cal Z} = \partial_z {\cal Z} + {\cal Z}\partial_z{\cal K}\,, \nonumber\\
&& \bar {\cal Z} = g_I\bar Z^I \, , \qquad
\overline{{\cal D}_z {\cal Z}} = \partial_{\bar z} \bar{\cal Z} + \bar{\cal Z}\partial_{\bar z}{\cal K}\,.
\end{eqnarray}
The potential may similarly be written as
\be\label{VZ}
V= 4\left({\bf n}\cdot {\bf n}\right)\rme^{\cal K}\left[ g^{z\bar z} {\cal D}_z{\cal Z}\,\overline{{\cal D}_z {\cal Z}}
 -3{\cal Z} \bar {\cal Z}  \right]\, .
\ee

\subsection{K{\"a}hler-gauge invariant formulation}

This potential (\ref{VZ}) is invariant under the  K{\"a}hler gauge transformations
\be
{\cal K}\to {\cal K} -\left(f+\bar f\right) \, , \qquad Z^I \to \rme^f Z^I\, ,\qquad \bar Z^I \to \rme^{\bar f} \bar Z^I\, ,
\ee
which induces the transformations
\be
A_\mu \to A_\mu + \ft12\rmi \partial_\mu\left(f-\bar f\right)\, , \qquad
{\cal Z} \to \rme^f {\cal Z}\, , \qquad
\bar {\cal Z} \to \rme^{\bar f} \bar{\cal Z}\, .
\ee
This suggests that we introduce the new, gauge-equivalent, K{\"a}hler
potential
\be
\tilde {\cal K} = {\cal K} + \log\left({\cal Z}\bar{\cal Z}\right)\, ,
\ee
and its associated, gauge-equivalent, K{\"a}hler connection,
\be
\tilde A_\mu = A_\mu  -\ft12\rmi \partial_\mu \log \left({\cal Z}/\bar {\cal Z}\right)\, .
\ee
In terms of the  function \cite{Behrndt:2000tr}
\be
W = \rme^{\tilde{\cal K}/2}\, ,
\ee
the scalar potential takes the manifestly K{\"a}hler-gauge invariant form
\be\label{Kinv}
V= 16\left({\bf n}\cdot{\bf n}\right) \left[ g^{z \bar z}\partial_z W \,\partial _{\bar z}W - \frac{3}{4} W^2\right] \, .
\ee
The supersymmetry transformation laws may be similarly written in K{\"a}hler-gauge invariant form
by introducing the new spinor parameters
\be\label{tildespin}
\tilde\e_i = \left({\cal Z}/\bar{\cal Z}\right)^{\frac{1}{4}} \e_i\, , \qquad
\tilde\e^i = \left(\bar {\cal Z}/{\cal Z}\right)^{\frac{1}{4}} \e^i\, .
\ee
One then finds that
\begin{eqnarray}\label{susytrans2}
\left({\cal Z}/\bar {\cal Z}\right)^{\frac{1}{4}}\delta\psi_{\m i}
&=&  \left(D_\mu -\ft12\rmi  \tilde A_\mu \right) \tilde \epsilon_i +
W  \gamma_\mu
\left[\tau_2\left({\bf n}\cdot\bftau\right)\right]_{ij}\tilde \epsilon^j \, , \nonumber\\
\left(\bar{\cal Z}/{\cal Z}\right)^{\frac{1}{4}}\delta\psi_\m^i
&=& \left(D_\mu +\ft12\rmi  \tilde A_\mu \right) \tilde\epsilon^i +
W\gamma_\mu
\left[\left({\bf n}\cdot\bftau\right)\tau_2\right]^{ij}\tilde \epsilon_j \, , \nonumber\\
  \left(\bar{\cal Z}/{\cal Z}\right)^{\frac{1}{4}}  \delta \lambda^ i
 &=& \slashed{\partial }\bar z\, \tilde\epsilon^i -4g^{z\bar z}\partial_z  W
\left[\left({\bf n}\cdot\bftau\right)\tau_2\right]^{ij}\tilde\epsilon_j \, , \nonumber\\
\left({\cal Z}/\bar {\cal Z}\right)^{\frac{1}{4}}\delta\lambda_i
 &=& \slashed{\partial }z\, \tilde\epsilon_i -4g^{z\bar z} \partial_{\bar z} W
\left[\tau_2\left({\bf n}\cdot\bftau\right)\right]_{ij}\tilde\epsilon^j \, .
\end{eqnarray}

\subsection{The model}

We will choose the prepotential
\begin{equation}
  F(Z)=\frac{\rmi}{4}(-Z^0Z^0+Z^1Z^1)\, ,
 \label{FXmin}
\end{equation}
which yields
\be\label{KgcalN}
\rme^{-{\cal K}(z,\bar z)}=Z^0\bar Z^0 - Z^1\bar Z^1\, .
\ee
Because of the homogeneity of $F$, we may choose $Z^0=1$, and we may then choose a parametrization such that $Z^1=z$.  Thus, without loss of generality we may choose
\begin{equation}
  Z^I=(1,z)\, .
 \label{Z0Z1}
\end{equation}
This yields
\be\label{KgcalN2}
{\cal K}(z,\bar z)= -\log\left(1- z\bar z\right)\, , \qquad {\cal Z} = g_0 + g_1 z\, ,
\ee
and hence
\be
   g_{z\bar z} =\left( 1- z\bar z\right)^{-2}\, , \qquad
    A_\mu =
 -\ft12\rmi  \left(1-z\bar z\right)^{-1}\left[ \bar z\partial_\mu z - z\partial_\mu \bar z\right]\, ,
\ee
and
\be
W^2 = \frac{\left(g_0+g_1 z\right)\left(g_0+g_1 \bar z\right)}{1-z\bar z}\,  .
\ee
The values of the complex scalar field $z$ must be restricted to the unit disc $z\bar z<1$, and the
metric is then the $\SU(1,1)$-invariant hyperbolic metric on this disc. From the formula (\ref{Kinv})
we find that
\be
V= \frac{4\left({\bf n}\cdot{\bf n}\right)}{1-z\bar z}\,  \left[ g_0^2\left(z\bar z-3\right) -2g_0 g_1\left(  z  + \bar z\right) +
g_1^2\left(1-3z\bar z\right)\right]\, .
\ee
When $g_0=g_1$ there is no extremum of $V$ within the unit disc, $z\bar z<1$.
Otherwise there is an extremum, which is at $z=0$ for the two special cases in which either $g_0=0$ or $g_1=0$.

As we will see in the section to follow, standard reality conditions imply that ${\bf n}$ is a real 3-vector, in which case $V>0$ when $g_0=0$ (but $g_1\ne0$), and the minimum of $V$ at $z=0$ is a supersymmetry breaking dS vacuum. In contrast, $V<0$ when $g_1=0$ (but $g_0\ne0$), and the maximum at $z=0$ is a supersymmetric adS vacuum. This is the case that we will focus on in this paper.  For our purposes, it will suffice to consider $g_I=(1,0)$,  so that
\be
{\cal Z}=1\, , \qquad W=  1/\sqrt{1-z\bar z}\, .
\ee
The scalar potential for this model is
\be\label{simpot}
V= 4 \left({\bf n} \cdot{\bf n}\right) \left[ \frac{z\bar z-3}{1-z\bar z}\right]\, .
\ee
For ${\bf n}$ a real 3-vector, this is precisely the potential of the $\SO(4)$ gauged $N=4$ supergravity, which has an identical scalar field content. It follows that we are considering an $N=2$ truncation of this $N=4$ model, and hence of $SO(8)$ gauged $N=8$ supergravity, as also follows from results of
 \cite{Duff:1999gh} on the $\U(1)^4$ truncation of the $N=8$ theory\footnote{This involves keeping only the 35 scalars parametrizing $\Sl(8;R)/\SO(8)$, as a first step, then using the local $\SO(8)$ invariance to diagonalize  the $\Sl(8;R)$ matrix, and retaining only the $8$ diagonal entries $X_\alpha$ ($\alpha=1,\dots,8$), as a second step. The potential depends only on the 7 scalars parametrizing the  subspace defined by $\prod_\alpha X_\alpha=1$, and a further truncation obtained by choosing the particular solution $X_1=X_2=X_3=X_4=X$ and $X_5=X_6=X_7=X_8=X^{-1}$ yields our model with $X=\rme^{\sigma/2}$.}. As the $N=8$ theory is a consistent truncation of the  $S^7$-compactification of
 $D=11$ supergravity, any domain-wall solution of our model, such as the one found later, can be lifted to $D=11$, using e.g. the results of \cite{Cvetic:1999xx,Cvetic:1999au}.

Our next task is to consider the implications for this model of  `twisted' reality conditions on the
fermion fields.

\subsection{Reality conditions}

We use conventions for which the charge conjugation matrix is
$\g^0$, so that all Dirac matrices $\g_\m$ are real and  $\gamma
_5=\rmi\gamma _0\gamma _1\gamma _2\gamma _3$ is imaginary. This implies
that complex conjugation changes the chirality of the spinors. In our
conventions, complex conjugation does not change the order of spinors, so
$(\lambda \chi )^*=\lambda ^*\chi ^*$. Although chiral spinors are
necessarily complex, each component of a chiral spinor must be linearly
related to the complex conjugate of the corresponding anti-chiral spinor
in order that each spinor has 4 independent complex components as
required by $N=2$ supersymmetry. In other words, we require a `reality
condition' of the form \cite{Pilch:1984aw,deWit:1987sn,Bergshoeff:2000qu}
\be\label{twist}
\epsilon^i = M^i{}_j\left(\epsilon_j\right)^*\,,
\ee
and similarly for other spinors, where $M^i{}_j$ are the entries of an invertible matrix $M$ that must be hermitian for reality of the action.
In particular, the redefined spinor parameters $\tilde \e^i$ and $\tilde\e_i$ introduced in (\ref{tildespin}) will satisfy exactly the same reality condition as the spinors $\e^i$ and $\e_i$.
By a redefinition of the spinors one can send
\be
M \to S^\dagger M\, S\,,
\ee
where $S$ is any invertible matrix, and one may choose $S$ so
as to bring $M$ to one of two standard
forms: $M=1$ or $M={\bf m}\cdot \bftau$ for some real unit 3-vector ${\bf m}$. The choice $M=1$ leads to
standard $N=2$ supergravity while the choice
$M={\bf m}\cdot\bftau$ (e.g. $M=\tau_3$) leads to $N=2$
pseudo-supergravity. In the former  case
the spinors are $\SU(2)$ doublets and in the latter case they are $\SU(1,1)$ doublets.

Consistency of the `reality condition' (\ref{twist}) with the fermionic field supersymmetry variations of
(\ref{susytrans1}) requires (for $g_I$ real; we henceforth restrict to
$g_0=1$ and $g_1=0$)
\be\label{consist}
\left({\bf n}\cdot\tau\right)\tau_2 =  M \left({\bf n}^*\cdot\tau\right)\tau_2 M^*\, .
\ee
The fermion field variations of (\ref{susytrans2})  may now be written, suppressing
$\SU(2)$ or $\SU(1,1)$ indices, in terms of a complex chiral doublet photino field $\lambda$ and a complex  anti-chiral doublet gravitino field $\psi_\mu$.
Recalling that ${\cal Z}=1$ for the model of interest here, we have
\begin{eqnarray}
\delta\psi_\mu & =  & \left( D_\mu - \ft12\rmi  A_\mu \right) \epsilon
 +  W \gamma_\mu \tau_2 \left(\bftau\cdot {\bf n}\right) M  \epsilon^*\,, \nonumber \\
\delta\lambda &=& \slashed{\partial }z\,  \epsilon -
4 g^{z\bar z}\partial_{\bar z} W\, \tau_2\left({\bf n}\cdot\bftau\right) M \epsilon^* \, ,
 \label{simpletransf2}
\end{eqnarray}
where $\epsilon$ is a complex anti-chiral doublet spinor parameter.

The consistency condition (\ref{consist}) should be viewed as a reality condition on the 3-vector ${\bf n}$. For $M=1$ it implies that ${\bf n}$ is real,
whereas for $M={\bf m}\cdot\bftau$, it implies that
the components of ${\bf n}$ perpendicular to ${\bf m}$ are real but the component (anti)parallel to ${\bf m}$ is imaginary. In either case we may write
\be
{\bf n} = g \, {\bf m} + {\bf n}_\perp \, , \qquad {\bf n}_\perp \cdot {\bf m}=0\, ,\qquad {\bf
n}_\perp,\,{\bf m}\in \mathbb{R}^3\,,
\ee
where $g$ is real\footnote{This is the coupling constant mentioned in the Introduction;
it can be viewed as the $\U(1)$ gauge coupling constant because it follows from (\ref{gPI})
that the vector coupling constant is really $g g_I$ and we have set $g_I=(1,0)$.} for $M=1$ (in which case ${\bf m}$ should be interpreted as an arbitrary unit 3-vector) but  imaginary \cite{deWit:1987sn} for $M={\bf m}\cdot\bftau$. In the former case we have $V<0$ and the potential
has a supersymmetric adS maximum at $z=0$.  In the latter case, the potential may be positive, negative or zero, depending on the relative magnitudes of $g$ and ${\bf n}_\perp$. In particular, when
$g=0$, but ${\bf n}_\perp \ne {\bf 0}$, we again have $V<0$ with a supersymmetric adS vacuum at $z=0$,
but the isometry supergroup is $\OSp(1,1|4)$ rather than  $\OSp(2|4)$. When ${\bf n}_\perp={\bf 0}$ but $g\ne0$ we have $V>0$
and what was the supersymmetric adS vacuum is now a supersymmetric dS vacuum, with isometry supergroup $\OSp(2^*|2,2)$.
The various possible supersymmetric (a)dS vacua, along with their isometry supergroups (and bosonic subgroups) are shown in table  \ref{tbl:d4N2dSAdS}.
\begin{table}[ht]
  \caption{\it (Pseudo)supersymmetric (a)dS vacua for $D=4$, $N=2$ supergravity
minimally coupled to one vector multiplet.}\label{tbl:d4N2dSAdS}
\begin{center}
\begin{tabular}{|l|c|c|}
\hline
Supergravity  & ${\bf n} \perp {\bf m}$  & ${\bf n} \parallel {\bf m}$       \\
\hline
Normal    & \multicolumn{2}{c|}{adS$\,\times \SO(2)$}   \\
$M=1$
&\multicolumn{2}{c|}{$\OSp(2|4)$}   \\
\hline
Pseudo  & adS$\,\times \SO(1,1)$ & dS$\,\times \SO(2)$ \\
$M={\bf m}\cdot\bftau$ & $\OSp(1,1|4)$ & $\OSp(2^*|2,2)$        \\
\hline
\end{tabular}
\end{center}
\end{table}

Henceforth, we make the standard choice for twisted reality conditions: $M=\tau_3$, corresponding to ${\bf m}=(0,0,1)$. We will also choose
\be
{\bf n}= g\, {\bf m} = g\, (0,0,1)\,.
\ee
This implies no loss of generality when $M=1$ but amounts to the choice ${\bf n}_\perp=0$
for  twisted reality conditions. In either case the potential (\ref{simpot}) becomes
\be\label{potadS}
V= -4g^2\left[ \frac{3-|z|^2}{1-|z|^2}\right] \, ,
\ee
but $g$ is real for $M=1$ and imaginary for $M=\tau_3$.

Our interest in the supersymmetry transformation laws is primarily due to the fact that one gets the conditions for preservation of supersymmetry, in a bosonic background, by setting to zero the supersymmetry variations of the fermion fields. So, for simplicity, we  now set $\delta\psi_\m$ and $\delta\lambda$ to zero in  (\ref{simpletransf2}) to arrive at the supersymmetry preservation conditions
\begin{eqnarray}
0 & =  & \left( D_\mu - \ft12\rmi   A_\mu \right) \epsilon
 + \rmi g\, W \gamma_\mu \tau_1 M  \epsilon^*\,, \nonumber \\
0&=& \slashed{\partial }z\,  \epsilon -
4 \rmi g \, g^{z\bar z}\partial_{\bar z} W \tau_1M \epsilon^* \, ,
 \label{suspres}
 \end{eqnarray}
where  $W= 1/\sqrt{1-|z|^2}$ and either $M=1$ or $M=\tau_3$, with $g$ real for $M=1$ and imaginary for $M=\tau_3$.

\subsection{Relation to fake (pseudo)supergravity}

For the models discussed in the previous section, the potential $V(z,\bar z)$ actually depends only on $|z|$. This suggests that we write
 \be
z= \rho(\sigma) e^{\rmi\phi} \, , \qquad \rho(\sigma) = \tanh\left(\sigma/2\right)\, ,
\ee
for real fields $\sigma$ and $\phi$. In terms of these new fields, the Lagrangian density is given by
\be\label{polarlag}
e^{-1}{\cal L} = \frac{1}{2} R - \frac{1}{4} \left[\left(\partial\sigma\right)^2 + \sinh^2\sigma \left(\partial\phi\right)^2 \right]  -V\, .
\ee
Moreover, we have
\be\label{WWprime}
W = \cosh\left(\sigma/2\right)\, \qquad W' \equiv \partial_\sigma W = \frac{1}{2}\sinh \left(\sigma/2\right)\, ,
\ee
and the potential is
\be\label{fakeV}
V= 16g^2 \left[\left(W'\right)^2 -\frac{3}{4}W^2\right] =  -4g^2\left(2+\cosh\sigma\right)\, .
\ee
The dependence of $V$ on $W$ is precisely that of fake (pseudo)supergravity.

The supersymmetry preservation conditions (\ref{suspres}) may be simplified by
writing the anti-chiral doublet of spinors $\e$ as
\be
\e = \frac{1}{2}\left(1-\gamma_5\right) \chi \, ,
\ee
where $\chi$ is a doublet of spinors satisfying the reality condition\footnote{This condition is consistent since $MM^*=1$ for both $M=1$ and $M=\tau_3$.}
\be
\chi^* = M\chi\, .
\ee
Passing to the new target space coordinates, we find that equations (\ref{suspres}) become
\begin{eqnarray}\label{suspres2}
0 &=&  \left[ D_\mu + \ft12\rmi  \gamma_5 A_\mu + gW\Gamma_\mu \tau_1 \right]\chi\,, \nonumber\\
0 &=& \left[ \Gamma^\mu\left(\partial_\mu\sigma + \rmi\gamma_5 \sinh\sigma\, \partial_\mu\phi\right)
-8gW' \tau_1\right]\chi\, ,
\end{eqnarray}
where
\be
A_\mu = \sinh^2\left(\sigma/2\right) \partial_\mu\phi \, ,
\ee
and
\be
\Gamma_\mu = \rmi\gamma_\mu\gamma_5\, ,
\ee
are alternative Dirac matrices\footnote{The matrices $\Gamma^0\Gamma_\mu$
are real and symmetric, so one may choose $\Gamma^0=\rmi \gamma^0\gamma_5$ as the charge conjugation matrix in this representation, replacing the choice $\gamma^0$ in the representation $\gamma_\mu$.}.

The first of  equations (\ref{suspres2})  is the Killing spinor equation;
note that the field $\phi$ enters only through the K{\"a}hler connection
$A_\mu$, which is proportional to $\partial_\mu\phi$. For any
configuration that depends on a single coordinate, such as the  domain
wall and cosmological configurations of interest, the K{\"a}hler field
strength (which is the pullback of the K{\"a}hler 2-form) will vanish. This
means that the term involving $\phi$ in the Killing spinor equation is
irrelevant to the integrability conditions of this equation, which are
the same as those of the simpler equation \be\label{simpleKil}
\left(D_\mu + g\,  \Gamma_\mu  W\tau_1\right)\chi=0 \, . \ee This
equation is precisely of the fake-supergravity form (\ref{KSflat})  with
${\bf g}=(g,0,0)$. For a domain-wall or cosmology background of the type
to be considered here, it is known
\cite{Freedman:2003ax,Sonner:2005sj,Skenderis:2006jq} that this Killing
spinor equation implies that \be\label{dwckil} \left[\Gamma^\mu \partial
_\mu \sigma  - 8 g\,  W'  \, \tau_1\right]\chi=0\, . \ee This agrees with
the second of equations (\ref{suspres2}) if and only if $\phi$ is
constant, and it then follows that $A_\mu$ vanishes, so that
(pseudo)Killing spinors are in fact solutions of (\ref{simpleKil}).

To summarize, we have shown that necessary and sufficient conditions for a domain wall (cosmology) to  be a supersymmetric solution of our (pseudo)supergravity model are (i)  that $z$ have constant phase and (ii) that (\ref{simpleKil}) admit a non-zero solution for $\chi$.

\section{Domain-wall/Cosmology correspondence}
 \label{ss:DWCcorr}

The metric for a $D=4$ flat domain-wall spacetime may be put in the standard form
\be
ds^2= dr^2 + \rme^{2A(r)} ds^2({\rm Mink3})\, ,
\ee
where $r$ represents distance in a direction perpendicular to the wall, so that the geometry is determined by the function
$A(r)$, and Mink3 is the 3-dimensional Minkowski metric.
The generic isometries of this metric are those of the $D=3$ Poincar{\'e} group, and to preserve this symmetry we must take the scalar fields to depend only on
$r$. Given a solution of this form for a model with scalar field potential $V$, the DW/C correspondence states  that the same model but with scalar field potential $-V$ has a cosmological solution with metric
\be
ds^2 = -dt^2 + \rme^{2A(t)} dl^2({\rm E3})
\ee
and scalar fields that depend only on time, where E3 is a 3-dimensional Euclidean metric. This is of standard FLRW form with $A(t)$ being the logarithm of the scale factor.

We have seen that (pseudo)supersymmetric solutions of our model are such that the complex field $z$ has constant phase $\phi$. The $\phi=0$ truncation of (\ref{polarlag}) is clearly consistent, so we effectively have a model with a single scalar field $\sigma$. From the results of  \cite{Skenderis:2006jq}, we then learn that (pseudo)supersymmetric domain walls (cosmologies) are such that (in current notation and conventions)
\begin{eqnarray}\label{BPS}
\dot A &=& \pm  2 |g| W = \pm 2 |g|\cosh\left(\sigma/2\right)\, , \nonumber\\
\dot \sigma &=& \mp 8|g| W' = \mp 4|g|\sinh \left(\sigma/2\right)\, ,
\end{eqnarray}
where $W$ is the scalar superpotential appearing in  (\ref{simpleKil})-(\ref{dwckil}), given by (\ref{WWprime})
for the model in hand, and the overdot indicates differentiation with respect to the independent variable, which is the distance variable $r$ in the domain-wall case and the time variable $t$ in the cosmology case.

\subsection{Domain walls}

As a check of  equations (\ref{BPS}) for the domain-wall case let us return to  (\ref{suspres2}). The coupling constant $g$ is real, and we may assume it to be  positive without loss of generality.
If one assumes\footnote{This assumption is valid generically, but for the special case of the adS vacuum there will be additional Killing spinors for which this assumption is not valid; their existence ensures that
all supersymmetries are preserved in this adS vacuum.} that $\chi$ depends only on $r$ then the projection of the first of eqs. (\ref{suspres2}) in any direction parallel to the wall yields
\be\label{parallelproj}
\left(\dot A + 2gW\Gamma\right)\chi=0 \, , \qquad \Gamma \equiv \Gamma_r \tau_1\, ,
\ee
where the first term comes from the spin connection; note that there is no contribution from
the K{\"a}hler connection because its only non-zero component is  $A_r$ (as a consequence of the fact that $z$ is a function only of $r$).  Writing $\chi$ as the sum of eigenspinors of $\Gamma$,
\be
\chi = \chi_+ + \chi_- \, , \qquad \Gamma\chi_\pm = \pm \chi_\pm\, ,
\ee
we see that
\be
\left(\dot A + 2gW\right)\chi_+ + \left(\dot A -2gW\right)\chi_- =0\, .
\ee
Acting on this equation with $\Gamma$ yields
\be
\left(\dot A + 2gW\right)\chi_+ - \left(\dot A -2gW\right)\chi_- =0\, ,
\ee
and hence
\be\label{BPSA}
\left(\dot A \pm 2gW\right)\chi_\pm =0\, ,
\ee
for either choice of sign. Given $gW\ne0$ and $\chi\ne0$, it follows that
\be
\dot A = \pm 2gW\, ,\qquad \chi= \chi_\mp\, ,
\ee
for {\it either} the top sign {\it or} the bottom sign.  Given this restriction on $\chi$, the second of
equations (\ref{suspres2}) in a domain-wall background becomes
\be
\left(\dot\sigma + \rmi\gamma_5\sinh\sigma\, \dot \phi \pm 8gW'\right)\chi_\mp=0\, .
\ee
Multiplying this equation by $\Gamma$ and using the fact that $\Gamma$ anticommutes with $\gamma_5$, we get the equation
\be
\left(\dot\sigma - \rmi\gamma_5\sinh\sigma\, \dot \phi \pm 8gW'\right)\chi_\mp=0\, ,
\ee
and hence we deduce that
\be
\dot\phi=0\, , \qquad \dot\sigma = \mp 8gW'\, .
\ee
We have now confirmed both that $\dot\phi=0$ and the first-order equations (\ref{BPS}).

Returning now to the Killing spinor equation, we note that $A_r=0$ for $\dot\phi=0$, so
the component of this equation perpendicular to the wall is simply
\be\label{BPSchi}
\dot \chi_\mp = \pm gW\chi_\mp \, .
\ee
It follows that the Killing spinors  take the form
\be\label{explicitDW}
\chi = \rme^{A/2}\xi_\mp\, , \qquad \Gamma\, \xi_\mp = \mp \xi_\mp\, ,
\ee
for constant real spinor $\xi_\mp$. Note that because $\Gamma$ is real, the reality of $\xi_\mp$ is consistent with it being an eigenspinor of $\Gamma$.

For the top (bottom) sign in (\ref{BPS}) there is a solution only for $r>0$ ($r<0$). Let us choose the top sign, corresponding to $r>0$.  Positivity of $|z|\equiv\rho$ requires $\sigma>0$, and the solution compatible
with this requirement is
\be
\sigma = 2\log \coth \left(g r\right)\, ,
\ee
and hence
\be
\rho \equiv \tanh\left(\sigma/2\right) = {\rm sech}\left(2g r\right)\, , \qquad
\cosh\left(\sigma/2\right) = \coth\left(2g r\right)\, .
\ee
The equation for $A$ (again for the top sign)  is now easily solved, and the solution is
\be
A = \log \sinh\left(2g r\right)\, .
\ee
The domain wall metric is therefore
\be\label{DWmetric}
ds^2= dr^2 + \sinh^2\left(2g r\right)ds^2({\rm Mink3})\, .
\ee
This metric is singular at $r=0$ but is asymptotic to adS as $r\to\infty$. We therefore have a solution that
is defined for $r>0$ and is asymptotic to the adS vacuum with $\sigma\equiv0$ as $r\to\infty$.
The Killing spinors for this solution are
\be
\chi =  \left[\sinh\left(2g r\right)\right]^{\frac{1}{2}} \xi_- \, , \qquad
\tau_1 \Gamma_r \, \xi_-= -\xi_-\, ,
\ee
which shows that the domain-wall solution is half-supersymmetric. As noted earlier, our model is an $N=2$ truncation of $\SO(8)$ gauged $N=8$ supergravity, and any solution can be lifted to a solution of 11-dimensional supergravity. The domain wall solution found here can be shown to be equivalent to one found in  \cite{Cvetic:1999xx}, where the lift to $D=11$ was interpreted as a continuous distribution of M2-branes.

\subsection{Cosmologies}

As observed in \cite{Skenderis:2006jq}, we may interpret the first-order equations (\ref{BPS}) as equations determining a pseudo-supersymmetric cosmology; in this case $g$ is imaginary and we may choose
\be
g= \rmi |g|\, .
\ee
The cosmological counterpart of (\ref{parallelproj}) is then found to be
\be
\left(\dot A + 2|g|W\tilde\Gamma\right)\chi =0\, ,  \qquad \tilde\Gamma= \rmi\Gamma_0\tau_1\, .
\ee
If we now write $\chi= \chi_+ + \chi_-$ as in the domain-wall case, but now with
$ \tilde \Gamma\chi_\pm = \pm \chi_\pm$ then we  find as before that
\be
\dot A = \pm 2|g|W\, , \qquad \chi= \chi_\mp\, .
\ee
The `photino equation'  then leads, as before,  to
\be
\dot\phi=0\, , \qquad \dot\sigma = \mp8|g|W'\, .
\ee
We have now confirmed both that $\dot\phi=0$ and (\ref{BPS}) for the cosmology case.
The associated pseudo-Killing spinors are given by
\be
\chi = e^{A/2}\xi_\mp\, , \qquad \tilde\Gamma\, \xi_\mp = \mp \xi_\mp\, .
\ee
Although $\tilde\Gamma$ is imaginary, it anticommutes with $\tau_3$, so the projection onto an eigenspace of $\tilde\Gamma$ is compatible with the twisted reality condition
\be\label{twistxi}
\xi_\mp^* = \tau_3\, \xi_\mp\, .
\ee

Choosing the top sign in (\ref{BPS}), which now corresponds to a cosmological solution with $t>0$, we find that the pseudo-supersymmetric solution has
\be
\sigma= 2\log \coth\left(|g| t\right)\,,
\ee
and a metric
\be
ds^2= -dt^2 + \sinh^2\left(2|g| t\right) dl^2({\rm E3})\, .
\ee
There is a big bang singularity at $t=0$ after which we have an expanding universe that approaches the
pseudo-supersymmetric dS vacuum as $t\to\infty$. The (time-dependent) pseudo-Killing spinors
for this solution are
\be
\chi(t) = \left[\sinh\left(2 |g| t\right)\right] ^{\frac{1}{2}} \xi_- \, ,
\qquad \rmi\Gamma_0\tau_1\,  \xi_- = - \xi_-\, ,
\ee
where $\xi_-$ is a constant spinor subject to the twisted reality condition (\ref{twistxi}).

Just as our `standard'  supergravity model was the $N=2$ truncation of an $S^7$ compactification  of $D=11$ supergravity, our pseudo-supergravity model is an $N=2$ truncation of an adS$_7$ `compactification' of the (two-time) M*-theory \cite{Liu:2003qa}. It follows that the cosmological solution lifts to a (presumably supersymmetric) solution of M$^*$-theory, and it is possible  that its big bang singularity could be resolved in this context.

\section{Discussion}
  \label{ss:disc}
As originally conceived, pseudo-supersymmetry
\cite{Skenderis:2006jq,Skenderis:2006fb} was merely a property of
certain cosmological solutions, unrelated to any symmetry. In view of
the results of this paper we should perhaps refer to this original
concept as `fake' pseudo-supersymmetry, because we have seen that  it is
possible to view it as arising, in special cases, as a consequence of a
local `supersymmetry' of an underlying  `pseudo-supergravity'  theory in
much the same way as `fake  supersymmetry'  arises (again in special
cases) as a consequence of the local supersymmetry of some supergravity
theory.  Pseudo-supergravity theories have vector  ghosts, and are
therefore non-unitary, but their existence is still a non-trivial
mathematical fact, and the ghost  sector plays no role in our analysis.

The concept of pseudo-supersymmetric cosmology arose from an application
to fake supergravity of the domain-wall/cosmology (DW/C) correspondence,
which relates domain wall solutions of a model with a scalar potential
$V$ to cosmological solutions of the same model but with scalar potential
$-V$ \cite{Cvetic:1994ya,Skenderis:2006jq}. As a consequence one may view a given model
with potential $V$ as the `dual' of the same model with potential $-V$.
This extension of the correspondence from solutions to  models  is
trivial in the `fake' setting but non-trivial  in the supergravity
setting. We have shown that a particular $\U(1)$ gauged $N=2$, $D=4$
Maxwell-Einstein supergravity with an adS vacuum has a `dual'
pseudo-supergravity theory with a dS vacuum, found by imposing `twisted
reality'  conditions on the fermions. The two models are dual in the
sense that not only are the bosonic truncations the same up to the flip
of sign of the scalar potential, but also in the sense that a
supersymmetric domain wall of the standard supergravity theory is dual,
in the sense of the DW/C correspondence, to a supersymmetric cosmology of
the pseudo-supergravity theory. The (a)dS vacua of these theories are
special cases of this correspondence, with enhanced supersymmetry in both
cases.

One motivation for this work was to understand the implications of
pseudo-supersymmetry. One such implication was recently presented in
\cite{Chemissany:2007fg}, where it was shown that scaling solutions that
are pseudo-supersymmetric must describe geodesic curves in target space.
Another possible implication concerns stability in de Sitter space: the
DW/C correspondence maps the well-known Breitenlohner-Freedman (BF)
stability bound on scalar field masses in adS space to an upper bound on
scalar field masses in dS space, such that dS vacua can be
pseudo-supersymmetric only if all scalar field masses satisfy the bound
\cite{Skenderis:2006fb}.  It may be verified that this `cosmological'
bound {\it is} satisfied by the supersymmetric dS vacuum of the
pseudo-supergravity theories considered here; this is just a consequence
of the fact that the BF bound is satisfied by the supersymmetric adS
vacuum of the `standard'  theory.  An obvious question is  whether the
cosmological version of the BF bound may also be viewed as a stability
bound. In this context we note that results of a recent article
\cite{Bros:2006gs}  suggest that particles with masses above the bound
are indeed unstable, although FLRW spacetimes are known to be classically
stable for any mass; see \cite{math} for recent rigorous analysis of this
issue. In any case, the cosmological bound has a simple group theoretic
meaning when phrased in terms of particles created by scalar fields,
where a `particle'  in dS space is identified with a unitary irreducible
representation (UIR) of the de Sitter group. These UIRs are classified
into principal, complementary and discrete series. Particles with mass
above the bound correspond to  the principal series while particles with
positive mass below the bound correspond to the complementary series.  It
is a group theoretic fact that there are no unitary fermionic
complementary series \cite{group}, so that particles with a non-zero mass
below the bound cannot be paired with fermions by any symmetry acting on
a positive-definite Hilbert space. This shows that if
pseudo-supersymmetry is to be realized as a symmetry of a dS vacuum then
the Hilbert space on which this symmetry acts must be indefinite, as
indeed it is for the dS vacua of pseudo-supergravity theories because of
the vector ghosts.

Normal supersymmetry has been instrumental in many recent advances by
providing a means of obtaining otherwise inaccessible exact results, and
pseudo-supersymmetry could play a similarly important role in cosmology,
e.g. in the context of a dS/CFT correspondence. Our realization of the
DW/C correspondence in supergravity shows that pseudo-supersymmetry is
not an `accidental' property of cosmology but one that is related to a
genuine, and mathematically non-trivial, symmetry.

\section*{Acknowledgments.}

\noindent We are grateful to Pietro Fr{\`e}, who contributed to notes that
have been used for this paper. We would also like to thank U. Moschella
and I. Papadimitriou for discussions. KS is supported by NWO via the
Vernieuwingsimpuls grant ``Quantum gravity and particle physics''. P.K.T.
is supported by the EPSRC. P.K.T. and A.V.P. thank the Galileo Galilei
Institute for Theoretical Physics for hospitality and the INFN for
partial support. This work is also supported in part by the European
Community's Human Potential Programme under contract MRTN-CT-2004-005104
`Constituents, fundamental forces and symmetries of the universe'. The
work of A.V.P. is supported in part by the FWO - Vlaanderen, project
G.0235.05 and by the Federal Office for Scientific, Technical and
Cultural Affairs through the `Interuniversity Attraction Poles Programme
-- Belgian Science Policy' P6/11-P.


\providecommand{\href}[2]{#2}\begingroup\raggedright\endgroup


\begin{thebibliography}{10}

\bibitem{Freedman:2003ax}
  D.~Z.~Freedman, C.~N{\'u}\~nez, M.~Schnabl and K.~Skenderis,
  ``Fake supergravity and domain wall stability'',
  Phys.\ Rev.\ D {\bf 69}, 104027 (2004)
  [arXiv:hep-th/0312055].

\bibitem{Skenderis:1999mm}
K.~Skenderis and P.~K.~Townsend,
``Gravitational stability and renormalization-group flow,''
Phys.\ Lett.\ B {\bf 468}, 46 (1999) [arXiv:hep-th/9909070].

\bibitem{DeWolfe:1999cp}
O.~DeWolfe, D.~Z.~Freedman, S.~S.~Gubser and A.~Karch,
``Modeling the fifth dimension with scalars and gravity,''
Phys.\ Rev.\ D {\bf 62}, 046008 (2000) [arXiv:hep-th/9909134].

\bibitem{Celi:2004st}
  A.~Celi, A.~Ceresole, G.~Dall'Agata, A.~Van Proeyen and M.~Zagermann,
  ``On the fakeness of fake supergravity'',
  Phys.\ Rev.\ D {\bf 71} (2005) 045009
  [arXiv:hep-th/0410126].

\bibitem{Zagermann:2004ac}
  M.~Zagermann,
  ``N = 4 fake supergravity,''
  Phys.\ Rev.\  D {\bf 71} (2005) 125007
  [arXiv:hep-th/0412081].

\bibitem{Sonner:2007cp}
  J.~Sonner and P.~K.~Townsend,
  ``Axion-Dilaton domain walls and fake supergravity,''
  Class.\ Quant.\ Grav.\  {\bf 24} (2007) 3479
  [arXiv:hep-th/0703276].

\bibitem{Elvang:2007ba}
  H.~Elvang, D.~Z.~Freedman and H.~Liu,
  ``From fake supergravity to superstars,''
  arXiv:hep-th/0703201.


\bibitem{Sonner:2005sj}
  J.~Sonner and P.~K.~Townsend,
  ``Dilaton domain walls and dynamical systems'',
  Class.\ Quant.\ Grav.\  {\bf 23}, 441 (2006)
  [arXiv:hep-th/0510115].


\bibitem{Skenderis:2006jq}
  K.~Skenderis and P.~K.~Townsend,
  ``Hidden supersymmetry of domain walls and cosmologies,''
  Phys.\ Rev.\ Lett.\  {\bf 96}, 191301 (2006)
  [arXiv:hep-th/0602260];

\bibitem{Skenderis:2006rr}
K.~Skenderis and P.~K.~Townsend,
  ``Hamilton-Jacobi method for domain walls and cosmologies,''
  Phys.\ Rev.\  D {\bf 74}, 125008 (2006)
  [arXiv:hep-th/0609056].

\bibitem{Skenderis:2006fb}
  K.~Skenderis and P.~K.~Townsend,
  ``Pseudo-supersymmetry and the domain-wall / cosmology correspondence,''
  J. Phys. A: Math. Theor. {\bf 40} (2007) 6733-6741 [arXiv:hep-th/0610253].

\bibitem{Cvetic:1994ya}
  M.~Cvetic and H.~H.~Soleng,
  ``Naked singularities in dilatonic domain wall space times,''
  Phys.\ Rev.\  D {\bf 51} (1995) 5768
  [arXiv:hep-th/9411170].

\bibitem{Nahm:1977tg}
  W.~Nahm,
  ``Supersymmetries and their representations,''
  Nucl.\ Phys.\  B {\bf 135} (1978) 149.

\bibitem{Lukierski:1984}
J.~Lukierski and A.~Nowicki, ``Supersymmetry in the presence of positive
cosmological constant'', preprint Wroc{\l}aw, no. 609, March 1984


\bibitem{Pilch:1984aw}
  K.~Pilch, P.~van Nieuwenhuizen and M.~F.~Sohnius,
  ``De Sitter superalgebras and supergravity,''
  Commun.\ Math.\ Phys.\  {\bf 98} (1985) 105.

\bibitem{Lukierski:1984it}
  J.~Lukierski and A.~Nowicki,
  ``All possible de Sitter superalgebras and the presence of ghosts,''
  Phys.\ Lett.\  B {\bf 151} (1985) 382.

\bibitem{deWit:1987sn}
  B.~de Wit and A.~Zwartkruis,
  ``SU(2,2/1,1) supergravity and $N=2$ supersymmetry with arbitrary cosmological constant'',  Class.\ Quant.\ Grav.\  {\bf 4} (1987) L59.

\bibitem{D'Auria:2002fh}
  R.~D'Auria and S.~Vaul{\`a},
  ``$D = 6$, $N = 2$, $F(4)$-supergravity with supersymmetric de Sitter
  background,''
  JHEP {\bf 0209} (2002) 057
  [arXiv:hep-th/0203074].


\bibitem{Bergshoeff:2007cg}
  E.~A.~Bergshoeff, J.~Hartong, A.~Ploegh, J.~Rosseel and D.~Van~den~Bleeken,
  ``Pseudo-supersymmetry and a tale of alternate Realities,''
  arXiv:0704.3559 [hep-th].

\bibitem{deWit:1984pk}
B.~de~Wit and A.~Van~Proeyen,
``Potentials and symmetries of general gauged
  $N=2$ supergravity -- Yang-Mills models'',
Nucl. Phys. {\bf B245} (1984) 89

\bibitem{VanProeyen:1999ni}
  A.~Van Proeyen,
 ``Tools for supersymmetry'', Annals of the University
of Craiova, Physics AUC {\bf 9 (part I)} (1999) 1--48,
\href{http://www.arXiv.org/abs/hep-th/9910030}{{\tt hep-th/9910030}}


\bibitem{Cecotti:1985mx}
  S.~Cecotti, L.~Girardello and M.~Porrati,
  ``Ward identities of local supersymmetry and spontaneous breaking of extended
  supergravity,'' in {\it New trends in particle theory}, proc. of the 9th Johns Hopkins
  Workshop, Firenze, World scientific, 1985, ed. L. Lusanna

\bibitem{Behrndt:2000tr}
  K.~Behrndt and M.~Cveti\v{c},
  ``Anti-de Sitter vacua of gauged supergravities with 8 supercharges,''
  Phys.\ Rev.\  D {\bf 61} (2000) 101901
  [arXiv:hep-th/0001159].

\bibitem{Ferrara:1985gj}
  S.~Ferrara and L.~Maiani,
  ``An introduction to supersymmetry breaking in extended supergravity,''
  in {\it Relativity, supersymmetry and cosmology}, proc. of SILARG V, 5th
Latin American Symp. on Relativity and Gravitation, Bariloche, Argentina,
Jan 1985,
  World Scientific, ed. O. Bressan, M. Castagnino, V.H. Hamity


\bibitem{Cecotti:1984wn}
  S.~Cecotti, L.~Girardello and M.~Porrati,
  ``Constraints on partial superhiggs,''
  Nucl.\ Phys.\  B {\bf 268} (1986) 295.

\bibitem{Duff:1999gh}
  M.~J.~Duff and J.~T.~Liu,
 ``Anti-de Sitter black holes in gauged $N = 8$ supergravity,''
  Nucl.\ Phys.\  B {\bf 554}, 237 (1999)
 [arXiv:hep-th/9901149].

\bibitem{Cvetic:1999xx}
  M.~Cveti{\v c}, S.~S.~Gubser, H.~L{\"u} and C.~N.~Pope,
  ``Symmetric potentials of gauged supergravities in diverse dimensions and
  Coulomb branch of gauge theories,''
  Phys.\ Rev.\  D {\bf 62}, 086003 (2000)
  [arXiv:hep-th/9909121].


\bibitem{Cvetic:1999au}
  M.~Cveti\v{c}, H.~L{\"u} and C.~N.~Pope,
  ``Four-dimensional $N = 4$, $\SO(4)$ gauged supergravity from $D = 11$,''
  Nucl.\ Phys.\  B {\bf 574} (2000) 761
  [arXiv:hep-th/9910252].



\bibitem{Bergshoeff:2000qu}
E.~Bergshoeff and A.~Van~Proeyen, ``The many faces of OSp$(1|32)$'',
Class. Quant. Grav. {\bf 17} (2000) 3277--3304,
\href{http://www.arXiv.org/abs/hep-th/0003261}{{\tt hep-th/0003261}}


\bibitem{Liu:2003qa}
  J.~T.~Liu, W.~A.~Sabra and W.~Y.~Wen,
  ``Consistent reductions of IIB*/M* theory and de Sitter supergravity,''
  JHEP {\bf 0401}, 007 (2004)
  [arXiv:hep-th/0304253].


\bibitem{Chemissany:2007fg}
  W.~Chemissany, A.~Ploegh and T.~Van Riet,
  ``A note on scaling cosmologies, geodesic motion and pseudo-susy,''
  arXiv:0704.1653 [hep-th].

\bibitem{Bros:2006gs}
  J.~Bros, H.~Epstein and U.~Moschella,
  ``Lifetime of a massive particle in a de Sitter universe,''
  arXiv:hep-th/0612184.

  \bibitem{math} H. Ringstr{\"o}m, ``Future stability of the Einstein-Non-Linear
scalar field system'', preprint.


\bibitem{group}
R. Takahashi,
``Sur les repr{\'e}sentations unitaires des groupes de Lorentz g{\'e}n{\'e}ralis{\'e}s'',
Bulletin de la Soci{\'e}t{\'e} Math{\'e}matique de France, 91 (1963), p. 289-433






\end{thebibliography}
\end{document}